\documentclass[titlepage,twoside,12pt]{article}
\usepackage{amssymb}
\usepackage{amsfonts}
\begin{document}

{\LARGE \bf Further de-Entangling Entanglement} \\ \\


{\bf Elem\'{e}r E ~Rosinger} \\ \\
{\small \it Department of Mathematics \\ and Applied Mathematics} \\
{\small \it University of Pretoria} \\
{\small \it Pretoria} \\
{\small \it 0002 South Africa} \\
{\small \it eerosinger@hotmail.com} \\ \\

{\bf Abstract} \\

A recent general model of entanglement, [6], that goes much beyond the usual one based on
tensor products of vector spaces is further developed here. It is shown that the usual
Cartesian product can be seen as two extreme particular instances of non-entanglement. Also
the recent approach to entanglement in [9] is incorporated in the general model in [6]. The
idea pursued is that entanglement is by now far too important a phenomenon in Quantum
Mechanics, in order to be left confined to its present exclusive modelling by tensor products.
Once this is realized, it turns out that one can quite easily {\it de-entangle} entanglement
from tensor products, and in fact, one can do so in a large variety of ways. Within such
general settings the issue of entanglement becomes connected with the issue of "System versus
Subsystem" in General Systems Theory, where synthesizing subsystems into a system may often be
less difficult than identifying subsystems in a system. \\ \\ \\

{\bf 0. Preliminaries} \\

It is by now well and widely understood that entanglement is an essential physical phenomenon,
and in fact, also a major resource, in Quantum Mechanics. As such, entanglement appeared for
the first time in the celebrated 1935 EPR paper of Einstein, Podolsky and Rosen, [1]. The term
{\it entanglement} itself, however, was not used in that paper. Instead, it was introduced, in
German, as {\it Verschr\"{a}nkung}, in the subsequent papers of Schr\"{o}dinger, [3,4], in
which he commented on the state of Quantum Mechanics at the time, and among others, brought to
attention the problematic situation which ever since would be called {\it Schr\"{o}dinger's
cat}. \\
In this regard, it may be instructive to note how often in science fundamental ideas, concepts
or phenomena may take quite some time to reach their better appreciation and understanding.
Indeed, until recently, hardly any of the major texts on Quantum Mechanics found it necessary
to deal with entanglement as a subject on its own, let alone, an important one. \\

As it happened, independently, and prior to the EPR paper, in Multilinear Algebra, the concept
of {\it tensor product} was introduced by mathematicians in view of its {\it universal}
property of establishing a natural connection between multilinear and linear mappings, see
Appendix in [6]. And it took some time, both for physicists and mathematicians, to become
aware of the fact that a natural mathematical formulation of quantum entanglement can be
obtained with tensor products of Hilbert spaces. \\
It may be of a certain historical interest to find out the first publication in Quantum
Mechanics where entanglement was treated using tensor products. \\

The aim of [6] was as follows. So far, entanglement has been modelled mathematically only by
tensor products of vector, and in particular, Hilbert spaces. However, in view of the
significant importance of entanglement, one could ask the question :

\begin{itemize}

\item Can entanglement be modelled in other more {\it general} ways, than by tensor products
of vector spaces ?

\end{itemize}

In [6] an affirmative answer was given to that question, by presenting general, and yet simple
ways of entanglement, which contain as a particular case tensor products. In fact, in that
general approach to entanglement, and unlike in the particular case of tensor products of
vector spaces, the spaces involved can be rather arbitrary sets, just as in the case of
Cartesian products, thus in particular, they need {\it not} be vector spaces, and not even
groups or semigroups. This paper further develops the respective results. \\

Here however, it is important to note the following. The issue of entanglement - when
de-entangled from its present tensor product based exclusive representation - becomes clearly
connected with the well known and rather deep issue of the relationship "system versus
subsystem" in General Systems Theory. And this system-subsystem relationship has two {\it dual}
aspects. One is to {\it synthesize} given subsystems into a system, while the other is to {\it
identify} subsystems in a given system. And as is well known, typically neither of the two are
easy issues, with the latter being often considerably more difficult, than the former,
[9-11]. \\

An easy to grasp illustration of that {\it asymmetry} in difficulty can be given by the so
called "Universal Law of Unintended Effects" in human affairs, a law which operates, among
others, precisely due to the usual lack of appropriate insight into the structure of
subsystems in a given system. \\
By the way, in medicine, the so called side-effects of treatments are such typical unintended
effects. \\

As it happens, entanglement has so far only been considered in the context of {\it synthesis},
that is, when independent quantum systems $S_1, \ldots , S_n$, where $n \geq 2$, with the
corresponding Hilbert spaces $H_1, \ldots , H_n$, are constituted into an aggregate quantum
system $S$, with the resulting Hilbert space $H = H_1 \bigotimes \ldots \bigotimes H_n$. \\

Such a synthesis approach, in a significantly generalized manner, was also pursued in [6]. \\

In [9], the dual systems approach, that is, of {\it identifying} subsystems in a system, is
pursued to a good extent, even if the authors may insists, as their very title indicates, to
have gone beyond any systemic considerations, by introducing observers in the process of
instituting and identifying entanglement. Indeed, the introduction of observers "$O$" does not
do more than simply {\it extend} the initial quantum system "$S$" assumed to be without
observers, to a new system "$S$ and $O$" which this time contains both the quantum system and
the observers. Of course, one may miss that point, or simply refuse to have mixed together
into a whole the quantum system and the observers. However, the fact remains that, as for
instance in [9], the observers enter into a highly relevant interaction with the quantum
system, not least in the ways entanglement is instituted and then identified by them. \\
Therefore, the merit - and novelty - in [9] is precisely in the fact that their approach to
the issue of entanglement is no longer confined to the usual systems synthesis leading to a
given tensor product, thus to one and only way to have entanglement. \\
Instead, in [9], even if still tensor products are used exclusively in modelling entanglement,
the way entanglement is instituted and identified is due to the observers who have a certain
latitude in identifying subsystems in the given quantum system. \\

In the sequel when going significantly beyond the usual tensor products, we shall mainly
pursue the approach in [6] which is along the usual {\it systems synthesis}. What is somewhat
unexpected with such an approach is that it can help in a similarly general way beyond tensor
products, this time along the dual approach of {\it subsystems identification} as well which,
quite likely, is to be considerably more difficult in its fuller study. \\
In this regard, the {\it subsystems identification} approach related to entanglement suggested
in [9], namely, by the introduction of observers, can be seen as a {\it particular} case of
the approach in this paper. \\
The advantage of such a particularization is in the fact mentioned above, namely that, the
subsystems identification approach is typically far more difficult than that of systems
synthesis. \\

As for a general enough approach to entanglement along subsystems identification lines, this
may quite likely be a considerably difficult task. \\

Two further observations are important, before proceeding with the paper. \\

The usual, tensor product based concept of entanglement is in fact given by the {\it negation}
of a certain kind of rather simple representation. Consequently, any extension and/or
deepening of that concept is bound to open up a large variety of meaningful possibilities.
This is precisely one of the features - often overlooked - of the concept of entanglement
which makes it nontrivial. Indeed, the role in Physics of definitions by negation is an issue
which can touch upon fundamental aspects, [5]. \\

Quantum physics arguments expressing quite some concern related to the usual tensor product
based concept of entanglement were recently presented in [8]. And they indicate what may be
seen as a lack of {\it ontological robustness} of that concept. As an effect, one may expect
that what appears to be entanglement in terms of tensor products may in fact correspond to
considerably deeper and more general aspects. In this regard, the old saying that "the whole
is {\it more} than the sum of its parts" may in fact mean that what is involved in that "more"
in the case of entanglement can correspond to very different things, depending on the specific
situations. \\

A likely consequence of these two facts is that, when seen in its depth and generality, the
concept of entanglement may naturally and necessarily branch into a larger variety of rather
different concepts which are only somewhat loosely related to one another. \\

The main definition, [6,7], is presented in section 3, and it is further extended in section 5.
The main new results can be found in sections 4 and 5. \\ \\

{\bf 1. Generators and Bases} \\

For convenience, here and in the next two sections we recall a few concepts and results in [6,7].
These are within the usual, that is, {\it systems synthesis} approach to entanglement. \\

{\bf Definition 1.1.} \\

Given any set $X$, a mapping $\psi : {\cal P} ( X ) \longrightarrow {\cal P} ( X )$ will be
called a {\it generator}, if and only if \\

(1.1)~~~ $ \forall~~~ A \subseteq X ~:~ A \subseteq \psi ( A ) $ \\

and \\

(1.2)~~~ $ \forall~~~ A \subseteq A\,' \subseteq X ~:~ \psi ( A ) \subseteq \psi ( A\,' ) $ \\

Let us denote by \\

(1.3)~~~ $ {\cal G}en ( X) $ \\

the set of generators on $X$. \\

{\bf Examples 1.1.} \\

1) A trivial, yet as we shall see important, example of generator is given by $\psi =
id_{{\cal P} ( X )}$, that is, $\psi ( A ) = A$, for $A \subseteq X$. \\

2) Another example which is important in the sequel is obtained as follows. Given any binary
operation $\alpha : X \times X \longrightarrow X$, we call a subset $A \subseteq X$ to be
$\alpha$-{\it stable}, if and only if \\

(1.4)~~~ $ x, y \in A ~~\Longrightarrow~~ \alpha ( x, y ) \in A $ \\

Obviously, $X$ itself is $\alpha$-stable, and the intersection of any family of
$\alpha$-stable subsets is also $\alpha$-stable. Consequently, for every subset $A \subseteq
X$, we can define the smallest $\alpha$-stable subset which contains it, namely \\

(1.5)~~~ $ [ A ]_\alpha = \bigcap_{A \subseteq B,~ B ~\alpha-stable}~ B $ \\

Therefore, we can associate with $\alpha$ the mapping  $\psi_\alpha : {\cal P} ( X )
\longrightarrow {\cal P} ( X )$ defined by \\

(1.6)~~~ $ \psi_\alpha ( A ) = [ A ]_\alpha,~~~ A \subseteq X $ \\

which is obviously a generator. Furthermore, we have in view of (1.5) \\

(1.7)~~~ $ \forall~~~ A \subseteq X ~:~
       \psi_\alpha ( \psi_\alpha ( A ) ) = \psi_\alpha ( A ) $ \\

since as mentioned, $[ A ]_\alpha$ is $\alpha$-stable, and obviously $[ A ]_\alpha \subseteq
[ A ]_\alpha$. \\

We note that, in general, the relation $\psi ( \psi ( A ) ) = \psi ( A )$, with $A \subseteq
X$, need not hold for an arbitrary generator $\psi$. \\

3) A particular case of 2) above is the following. Let $( S,\ast )$ be a semigroup with the
neutral element $e$. Then $[ \{ e \} ]_\ast = \{ e \}$, while for $a \in S,~ a \neq e$, we
have $[ \{ a \} ]_\ast = \{ a, a \ast a, a \ast a \ast a, \dots \}$. \\

For instance, if $( S, \ast ) = ( \mathbb{N}, + )$, then $[ \{ 1 \} ]_+ = \mathbb{N} \setminus
\{ 0 \} = \mathbb{N}_+$. \\

4) A further case, which is of relevance in tensor products, is when we are given a vector
space $E$ over some field of scalars $\mathbb{K}$. If now we have any subset $A \subseteq E$,
then we can define $\psi ( A )$ as the vector subspace in $E$ generated by $A$. Clearly, we
obtain a generator in the sense of Definition 1.1. \\

Here one should, however, note that this generator is {\it no longer} of the simple form in 2)
above. Indeed, the generation of a vector subspace does involve {\it two} algebraic operations,
and not only one, as in 2) above, namely, addition of vectors, and multiplication of vectors
with scalars. \\

5) The general pattern corresponding to 4), and which contains 2) as a particular case is as
follows. Given on a set $X$ the mappings $\alpha_1, \dots , \alpha_n : X \longrightarrow X$
and $\beta_1 : K_1 \times X \longrightarrow X, \ldots , \beta_m : K_m \times X \longrightarrow
X$, where $K_1, \ldots , K_m$ are certain sets. Then we call a subset $A \subseteq X$ to be
$\alpha,\beta$-{\it stable}, if and only if, see (3.1) \\

(1.8)~~~ $ x, y \in A ~~\Longrightarrow~~ \alpha_i ( x, y ) \in A,~~~ 1 \leq i \leq n $ \\

and \\

(1.9)~~~ $ x \in A ~~\Longrightarrow~~ \beta_j ( c_j, x ) \in A,~~~
                                            c_j \in K_j,~~ 1 \leq j \leq m $ \\

Obviously, $X$ itself is $\alpha,\beta$-stable, and the intersection of any family of
$\alpha,\beta$-stable subsets is again $\alpha,\beta$-stable. Thus, for every subset $A
\subseteq X$, we can define the smallest $\alpha,\beta$-stable subset which contains it,
namely \\

(1.9)~~~ $ [ A ]_{\alpha,\beta} = \bigcap_{A \subseteq B,~ B ~\alpha,\beta-stable}~ B $ \\

Therefore, we can define the mapping  $\psi_{\alpha,\beta} : {\cal P} ( X ) \longrightarrow
{\cal P} ( X )$ as given by \\

(1.11)~~~ $ \psi_{\alpha,\beta} ( A ) = [ A ]_{\alpha,\beta},~~~ A \subseteq X $ \\

which is obviously a generator in the sense of Definition 1.1. Furthermore, we have in view of
(1.10) \\

(1.12)~~~ $ \forall~~~ A \subseteq X ~:~
       \psi_{\alpha,\beta} ( \psi_{\alpha,\beta} ( A ) ) = \psi_{\alpha,\beta} ( A ) $ \\

since as mentioned, $[ A ]_{\alpha,\beta}$ is $\alpha,\beta$-stable, and obviously
$[ A ]_{\alpha\beta} \subseteq [ A ]_{\alpha,\beta}$. \\

{\bf Definition 1.2.} \\

Given a generator $\psi : {\cal P} ( X ) \longrightarrow {\cal P} ( X )$, a subset $B
\subseteq X$ is called a $\psi$-{\it basis} for $X$, if and only if \\

(1.13)~~~ $ \psi ( B ) = X $ \\

and $B$ is minimal with that property, namely \\

(1.14)~~~ $ \forall~~~ B\,' \subsetneqq B ~:~ \psi ( B\,' ) \subsetneqq X $ \\

Let us denote by \\

(1.15)~~~ $ {\cal B}s_\psi ( X ) $ \\

the set of all $B \subseteq X$ which are a $\psi$-basis for $X$. \\

{\bf Note 1.1.} \\

1) In view of 3) in Examples 1.1., it follows that neither $\{ 0 \}$, nor $\{ 1 \}$ are
$\psi_+$-bases in $( \mathbb{N}, + )$, while on the other hand, $\{ 0, 1 \}$ is. \\

2) Within the setting in 3) in Examples 1.1., a subset $B \subset E$ is a $\psi$-basis in the
vector space $E$, if and only if it is a basis in the usual vector space sense. \\ \\

{\bf 2. Covered Generators} \\

The usual {\it systems synthesis} approach to entanglement is extended now considerably beyond
tensor products, based on the previous section 1. \\

{\bf Definition 2.1.} \\

Given the sets $X$ and $Y$, with the corresponding generators $\psi : {\cal P} ( X )
\longrightarrow {\cal P} ( X )$, $\varphi : {\cal P} ( Y ) \longrightarrow {\cal P} ( Y )$,
and $\chi : {\cal P} ( X \times Y ) \longrightarrow {\cal P} ( X \times Y )$. We say that $\chi$
is {\it covered} by $\psi, \varphi$, if and only if \\

(2.1)~~~ $ \forall~~~ A \subseteq X,~ B \subseteq Y ~:~
                    \chi ( A \times B ) \subseteq \psi ( A ) \times \varphi ( B ) $ \\

{\bf Examples 2.1.} \\

1) Obviously, if $\psi = id_{{\cal P} ( X )},~ \varphi = id_{{\cal P} ( Y )}$ and $\chi =
id_{{\cal P}( X \times Y )}$, then $\chi$ is a covering for $\psi, \varphi$. \\

2) Let now $\alpha : X \times X \longrightarrow X$ and $\beta : Y \times Y \longrightarrow Y$ be
two binary operations and, as usual, let us associate with them the binary operation $\alpha
\times \beta : ( X \times Y ) \times ( X \times Y ) \longrightarrow ( X \times Y )$ given
by \\

(2.2)~~~ $ ( \alpha \times \beta ) ( ( x, y ), ( u, v ) ) =
             ( \alpha ( x, u ), \beta ( y, v ) ),~~~ x, u \in X,~ y, v \in Y $ \\

Then $\psi_{\alpha \times \beta}$ is covered by $\psi_\alpha,~ \psi_\beta$, see [6], Lemma
2.1. \\

3) The case of interest for tensor products is the following. Given two vector spaces $E$ and
$F$ on a scalar field $\mathbb{K}$, let $\psi$ and $\varphi$ be the corresponding generators
as defined in 4) in Examples 1.1. Further, on the vector space $E \times F$, let $\chi$ be the
generator defined in the same manner. \\

Then it follows easily that $\chi$ is a covering for $\psi$ and $\varphi$. \\ \\

{\bf 3. A General Concept of Entanglement} \\

Within the usual {\it systems synthesis} approach, and based on the above, we can now give a very general definition of entanglement, [6, section 4, 7, section 3], much beyond that of the usual tensor tensor products, or in fact, of any algebraic nature as such. \\

{\bf Definition 3.1.} \\

Given the sets $X$ and $Y$, with the corresponding generators $\psi : {\cal P} ( X ) \longrightarrow {\cal P} ( X )$, $\varphi : {\cal P} ( Y ) \longrightarrow {\cal P} ( Y )$. \\

An element $w \subseteq X \bigotimes_{\psi, \varphi} Y$ is called {\it entangled}, if and only if it is {\it not} of the form \\

(3.1)~~~ $ w = x \bigotimes_{\psi, \varphi} y $ \\

for some $x \in X$ and $By \in Y$. \\

{\bf Note 3.1.} \\

1) In view of [7, Theorem 3.3.], the above Definition 3.1. contains as a particular case the customary concept of entanglement as formulated in terms of usual tensor products. \\

2) The interest in the general concept of entanglement in Definition 3.1. is, among others, in the fact that it is {\it no longer} confined within any kind of algebraic context. In this way, this paper, following [6,7], shows that entanglement can, so to say, be {\it de-entangled} not only from tensor products, but also more generally, from all algebra as well. \\

Consequently, the "quantum way of composition of system", namely, by the tensor product of their state spaces, is no 
longer limited to quantum systems alone. Indeed, in order to be composable by tensor products, such state spaces need no longer be limited to vector spaces, and instead, can now be given by arbitrary sets. \\

3) The generality of the concept of entanglement given in Definition 3.1., which at first appears confined to the systems synthesis approach, has nevertheless the further advantage to allow a certain incursion into the dual subsystems identification approach. \\ \\

\end{document}